\documentclass[preprint,showpacs,preprintnumbers,amsmath,amssymb,showkeys]{revtex4}

\usepackage{graphicx}
\usepackage{dcolumn}
\usepackage{bm}

\begin{document}

\preprint{APS/123-QED}

\title{Comparison of social and physical free \\energies on a toy model}

\author{Josip Kasac}%
\affiliation{Department of Robotics and Automation of
Manufacturing Systems, \\ Faculty of Mechanical Engineering and
Naval Architecture, \\University of Zagreb, Zagreb, Croatia}

\author{Hrvoje Stefancic}%
\affiliation{Theoretical Physics Division, \\ Rudjer Boskovic
Institute, \\ Zagreb, Croatia}

\author{Josip Stepanic}\altaffiliation{Corresponding author}\email{josip.j.stepanic@fsb.hr}
\affiliation{Department of Quality, \\ Faculty of Mechanical
Engineering and Naval Architecture,\\ University of Zagreb,
Zagreb, Croatia }%


\begin{abstract}
Social free energy has been recently introduced as a measure of
social action obtainable in a given social system, without changes
in its structure. The authors of this paper argue that social free
energy surpasses the gap between the verbally formulated value
sets of social systems and the quantitatively based predictions.
This point is further developed by analyzing the relation between
the social and the physical free energy. Generically, this is done
for a particular type of social dynamics. The extracted type of
social dynamics is one of many realistic types of the differing
proportion of social and economic elements. Numerically, this has
been done for a toy model of interacting agents. The values of the
social and physical free energies are, within the numerical
accuracy, equivalent in the class of non-trivial, quasi-stationary
model states.
\end{abstract}

\maketitle

\section{\label{sec:level1}Introduction }

The modelling of social systems based on notions from statistical
physics enriches the understanding of collective phenomena
[\onlinecite{M22} - \onlinecite{M24}]. Within that context the
social meaning of free energy has been explicitly addressed
[\onlinecite{M22} - \onlinecite{M5}]. Social free energy was
introduced as a measure of system resources which are unused in
regular, predicted functioning, but which are involved during
suppression of environmentally induced dynamics changes \cite{M5}.
Depending on the context, it was recognized as the combination of
innovation and conformity of a collective [\onlinecite{M22} -
\onlinecite{M24}], profit \cite{M1}, common benefit \cite{M2},
availability \cite{M3}, or free value of the canonical portfolio
\cite{M4}. The free energy in the references listed was introduced
at the quantitative level in the usual way (expression (\ref{eq7})
in this article) and was linked with its sociological
interpretation. The listed social interpretations of the physical
free energy imply existence of a socially relevant quantity,
analogue or at least closely related to physical free energy.
However, the very diversity of the notions and their independent
development show that a unified approach to recognizing social
meaning of the physical free energy is still missing.

This paper claims that social situations are interpreted (e.g.,
present situations described and future predicted) based on
evaluation of a social analogue of physical free energy.

The development of the social interpretation of the free energy is
by no means straightforward, as it invokes calculations in the
social, predominantly verbally formulated context. The following
scheme of the development is suggested: (i) contextualization of
the free energy, (ii) definition of the interacting agent toy
model in the context set, (iii) independent introduction of social
free energy and physical free energy within the model (the former
quantity is introduced from strictly qualitative, readily
recognizable considerations as the measure of particular social
action, while the latter is calculated using the well-founded
formalism of statistical mechanics, seemingly unrelated to the
social context of the model), (iv) calculation of these two
quantities for evolving state of the interacting agent toy model,
and a posteriori demonstration of the validity of the initial
claim for the model. This equality is the starting point for
further, more profiled analyses of relation between the two types
of free energies. Clear relation between the social and physical
free energy would provide one with a broadly applicable
quantitative mean for analysis of social systems' aggregated
quantities, which in turn contributes to better understanding of
the social system dynamics. In this sense we emphasize the
quantification of social context, as it has been rather vaguely
covered in the literature in comparison with the economic one. In
addition, we do not attempt to simplify the entire existing social
and economic dynamics to one simple model. Instead, among the
large number of existing types of social system dynamics we
concentrate on one particular type, apply the stated scheme onto
it, and subsequently model it. Still, we consider the defining of
social free energy, and the establishing of its relation with
physical free energy to be independent of the particular
represented type of the social system dynamics.

The paper is organized as follows. The social context of social
system dynamics is discussed in the second section in order to
elaborate the link between the general social system and its
projection on the quantifiable subsystem. The toy model
corresponding to a particular social behavior is described in the
third section. The model indicators are introduced in the fourth
section. Results of calculations of free energies are discussed in
the fifth section. The main results are summarized in the sixth
section.

\section{Basic elements of social system value set}

A value set is a qualitative structure attributed to a social
system, which collects formal (legislative) and informal (customs,
norms and values) rules governing complete social dynamics of a
social system. It is a fact of life that value sets differ
significantly. The interpretation of many types of transfer of
tradable goods, some of these being seemingly strictly economic,
requires the full value set of the corresponding social system.
Let us illustrate this by using the following two examples.

The first example is "the Melanesian culture of status-seeking
through gift giving. Making a large gift is a bid for social
dominance in everyday life in these societies, and rejecting the
gift is a rejection of being subordinate." [\onlinecite{M6},
p.159]. Gifts in that culture combine the economic context of what
is otherwise a valuable collection of resources with the social
one. In the ultimatum game experiment (described in detail by
Gintis \cite{M6}) in which participants pair-wisely arrange
transfer of resources they are initially given, on average
participants belonging to such a culture "offered more than half
pie, and many of these 'hyperfair' offers were rejected."
\cite{M6}. In contrast, the fair transfer within market trading
economy is considered to be the half pie.

The second example is the {\em potlatch}, the ritualized barter
ceremony often used to settle positions in communities of North
American Indians. "A person's prestige depended largely on his
power to influence others through impressive size of gifts
offered, and, since the debts carried interest, the 'giver' rose
in the eyes of the community to be \ldots  a person of
considerable standing." [\onlinecite{M7}a]. Valuable resources
were even destroyed in order to demonstrate the owner's wealth and
prestige [\onlinecite{M7}b].

These examples illustrate the point that the realistic collective
behavior incorporates a large number of types, some of which are
unrealistic if interpreted by different value sets. Taking these
diverse systems mutually on equal footing is useful in gaining
understanding of generic system quantities. Before proceeding, let
us stress these points, because a reduction of value sets, needed
for the sake of operationality, suppresses the social context and
leaves rather unrealistic behavior.

Despite the recognized importance of value sets in regulating
social dynamics, these constructs have been rarely linked in
detail. As an illustration, altruism and self-interest as two of
human characteristics are incorporated in diverse value sets with
different significance. However, their precise meaning is still
missing. Regarding this, recent literature points out that the
understanding of altruism is still changing significantly, which
includes the recognition of its sub-categories [\onlinecite{M8},
\onlinecite{M9}]. On the other hand, the boundary between
self-interest and altruism is questionable. The interpretation of
other human-related terms is similarly unsettled. All this
influences the interpretation of social dynamics and its
derivatives, e.g. simulation models.

As a consequence, for the sake of a definite interpretation of
simulation results, one needs to reduce the complexity of social
dynamics through its relation to quantifiable resources. The
reduction of social dynamics requires the reduction of the
corresponding value set. Reduction implies extracting the facts,
regarding observable actions which include resources, from the
value sets. The set of thus extracted facts does not belong to any
particular social system. Yet, its representative quality is
sufficient to justify its broadening and linking to a specific
system.

In this way, a prerequisite for determining the relation between
the social and physical free energy is formulated. Such a relation
contributes to simplifying the rules of social dynamics.
Furthermore, it contributes to the importance of existing
formalism of physics in the relatively new context.

The resources are quantifiable artifacts, objects, materials and
human characteristics (e.g. free time, skills, knowledge) linked
to social dynamics. The use of resources spans the range from
economic to social, as illustrated previously. Let us use the
following three types of resources, in which the proportion of
social context is prevalent or at least significant, to contribute
to the awareness of the importance of the socially-governed
transfers: (i) grants, writing off debts (which occurs from
individual to international level), money donations, etc., (ii)
donated blood, and (iii) socially responsible investments and
resources of charitable organizations (e.g. Salvation Army).

The processes including the listed types of resources share some
elements; e.g. the donated blood is a regularly observed gift
\cite{M10}. Blood donation
\begin{enumerate}
 \item has significant impact on the individuals and the whole collective,
 \item is strictly voluntary in the sense that there are no
  laws and penalties for potential blood donors who do not donate blood, and
 \item relies heavily on the presumed honesty and sincerity
  of the giver, despite the observed fallacies.\end{enumerate}

Therefore, blood donation itself is related presumably to the part
of the social system's value set which is separated from
economics. The extracted three points have been observed similarly
in the cases of other listed types of resources.

The common points in the transfer of different resource types can
be generalized: locally, there is non-homogeneous distribution of
resources. A part of the population experiences a lack, and the
other part a surplus of these. These conditions may be temporary
or permanent, adopted by the majority or a minority of population.
It is a fact that there are different processes which tend to
balance the non-homogeneity of resource distribution (along with
others which try to enhance the differences). These processes may
be complexly structured (supported loans which require proofs of
social status, and which are given in several time-separated
phases), they may be permanent (blood donation), triggered by some
event (help provided to people suffering from some natural
disaster), combined (donations), with or without institutions
mediating transfer of resources. These processes may be the
consequences of the fact that resources given to the people who
suffer from the lack of them eventually enable further collecting
of the resources from the people with current surplus, or the
consequences of socially responsible investments. They may be the
consequences of philanthropic character of individuals with
surplus of resources, or of their tendency to rise in the eyes of
local population, i.e. to make their social rank higher, and
augment the power which is related to the rank in the
corresponding social system. They are usually sensitive to some
characteristics of persons lacking resources, e.g., grants include
citizenship or age requirements, grants are given only to some
professions, help is given to neighbors, elderly, homeless, etc.
On the average, however, most of the population lacking resources
is eligible at least for some of the resource transfer processes.
Aside from that, resource transfers tend to be localized in
physical space, because the durability of resources,
administrative requirements, etc. raise cost of or otherwise
complicate the longer distance transfers. Moreover, there are
fewer types of resources shared by the dynamics of more distant
social systems.

As long as one is interested only in observable and quantifiable
part of the processes of the types mentioned, the value sets'
related points should be suppressed. Thus one ends with the
following rule expressing all relevant elements of the resource
transfer:

\begin{center}
{\em part of resources is transferred from people with surplus
\\ of resources to the neighboring people lacking resources}\end{center}

Before proceeding, the following point should be emphasized once
again: the rule stated is not appropriate for all the existing
resource transfers, e.g., for processes in market economies. It is
appropriate for situations in which social dimension of processes
is important, and expresses directly quantifiable part, which is
linked by the value set to the other, directly non-quantifiable
part. The exclusion of the value set includes the refraining from
interpretation about persons giving part of resources, i.e.,
whether they are altruists or self-interested.

\section{Model}

The model includes mutually interacting agents, their
configuration and environmental impact.

\begin{figure}
\includegraphics[width=4.7in, height=4.5in]{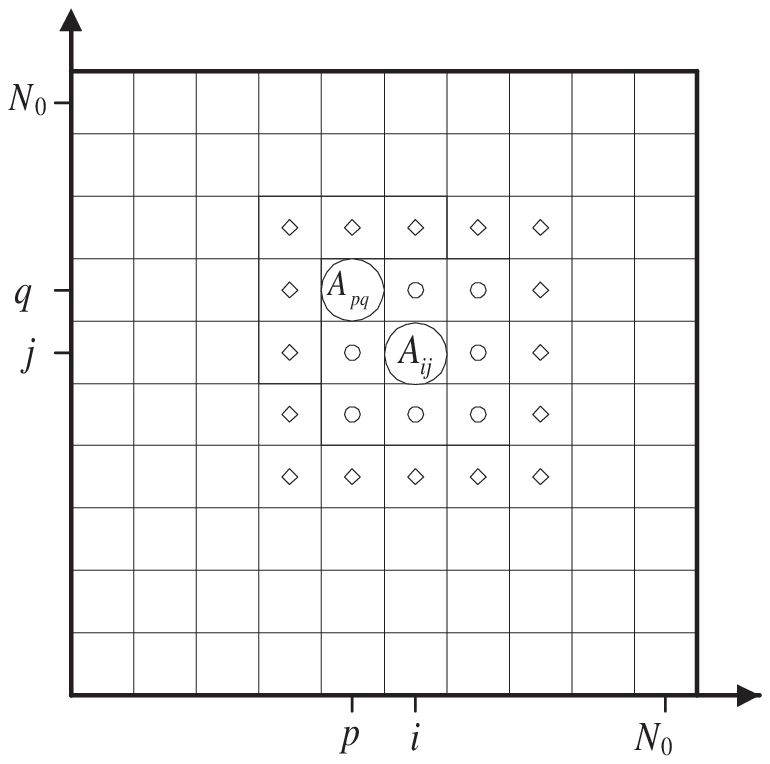}
\caption{\label{fig:f1} }
\end{figure}

Agents are fixed at nodes of two-dimensional net of dimensions
$N_0 \times N_0$, Fig.~\ref{fig:f1}. Coordinates of agent on the
$i$-th node in one direction and the $j$-th node in the other
direction are denoted as $m = (i, j)$. Such a net, generally
represents connections among agents. Hence it refers to
socio-biological, economical relations, or relations caused by
other interests among the agents. In some special cases it may
refer to physical space occupied by agents. This point is
addressed in more detail at the end of this section. Agents
collect the resource $u$, a scalar quantity, which is taken to be
a non-negative quantity. The amount of resources owned is a
positive number or zero. An agent with resources $u$ is considered
rich if $u  > u_0$, poor if $u_0 > u > 0$, and dead if $u = 0$. In
the context of the model the terms rich and poor refer to the
quantity of resources. In some special cases (donations, grants,
etc.) they coincide with their conventional economic meaning. If
resources of a particular agent become negative at some point of
time, they are set to zero and the agent is considered dead. Dead
agents are further excluded from the resource transfers. For a
rich agent with resources $u$ the difference $u - u_0$ is called
surplus of resources. Similarly, for a poor agent with resources
$u$, the difference $u_0 - u$ is called lack of resources. Let us
remark that our model can also describe the systems with different
types of resources if a fixed exchange ratio to a scalar quantity
$u$ is defined for each type of resources.

As a consequence of internal, otherwise unspecified dynamics,
agents regularly consume a finite value of resources $c$.

Because of the environmental influence, each agent's resources are
synchronously changed for a random amount $w(m)$. The distribution
of changes $w$ is the Gaussian distribution with a mean value $a$
and a variance $\mu$ . The mean value $a$ represents the average
resource change, and for a system we take $a > 0$. In each time
interval there are some resources obtained from the environment,
and some resources destroyed because of the influences from the
environment. If resources are smaller after the interaction with
the environment this means that destructive influences, e.g., fire
or flood, were stronger than the effects of making the resources
larger. The choice of a symmetrical function for distribution of
$w$ seemingly contradicts the usual skew shape of resource
distributions  \cite{M11}. However, since the positive (negative)
part of the Gaussian distribution represents the increasing
(reducing) the resources, it qualitatively collects the total
interaction of the agents with the environment.

Such a setup of the model includes the relevant agent
characteristics, in accordance with the definitions of the agent
[\onlinecite{M12}, \onlinecite{M13}], and the social agent
\cite{M14}. Each agent acts on himself or herself, which is taken
into account by the parameter $c$, and interacts with the
environment, which is included through $a$ and $\mu$. The agents
respond to the current environment state optimally in the sense
that all agents always follow all the system rules which,
nevertheless, does not assure them a sufficient amount of
resources. Formally, this assures the equality of the form of
distribution function for all agents. Owing to the simplicity of
the toy model, the agent and the environment characteristics are
somewhat mixed. On the one hand, the model covers the case of
agents of bounded rationality, which share knowledge about
environment, but the knowledge which does not include all the
rules underlying environment dynamics. On the other hand, the
agents in the model could have maximal possible knowledge about
environment dynamics, but in the environment which itself is
stochastic. In that sense, the optimal agent response means that
in the system there are no local fluctuations of knowledge, i.e.,
all agents have identical knowledge about the environment and the
processes of transfer of resources from the environment. In
addition, it is further assumed that agents know exactly the local
amount of resources. Further in the text the last assumption is
included into the rule of intra-agent resource transfer in which
resources of several neighboring agents are related.

The constancy of the parameters in space means that the system
latency and integrity are strong. The latency is taken here as a
collection of all modes the constant application of which enables
the agents to assure, preserve and reproduce both individual
motivation and cultural elements which generate and keep the
motivation. Integration here means a set of procedures regulating
the system components interaction. Furthermore, the adaptation
improves for larger $a$. For example, if the agents resemble
manufacturing firms, a better adaptation means a more intensive
consideration of customer needs and resource provider potentials -
clear signs of understanding of a part of environment complexity
\cite{M15}. Moreover, a better adaptation means that rapid changes
in $a$ are less probable.

Agents mutually interact through the transfer of resources in a
way described by the following algorithm, Fig.~\ref{fig:f1}: a
rich agent at location $m$ may give a part of his or her
resources, maximally the surplus $(u_m-u_0)$ to the neighboring
poor agents. Here, the neighboring are those for whom the indices
of position on the axes do not differ more than $1$. In accordance
with what has been stated before about the net, these may be the
agents closest in space but is not necessarily so. The rich agent
considers the total lack of resources of all his or her nearest
neighbor poor agents $\sum\limits_{i(m)} {(u_0  - u_i )}$. The
expression $i(m)$ means that all the poor agents at locations $i$
that are the nearest neighbors to the agent at location $m$, are
included. The rich agent divides the surplus among his or her poor
nearest neighbors. The amount of resources the agent could give to
the poor neighbor at location $n$ is $(u_m  - u_0 ) \cdot (u_0 -
u_n )/\sum\limits_{i(m)} {(u_0  - u_i )}$. Since a poor agent
could have several rich agents as its nearest neighbors, it
receives contributions from all of them. Their total surplus is
$\sum\limits_{j(n)} {(u_j  - u_0 )}$. Because of that, the
initially considered rich agent at location $m$ gives the
following part of the amount of resources:
\begin{equation}
(u_m  - u_0 ) \cdot \frac{{(u_0  - u_n )}}{{\sum\limits_{i(m)}
{(u_0  - u_i )} }} \cdot \frac{{(u_m  - u_0
)}}{{\sum\limits_{j(n)} {(u_j  - u_0 )} }}, \label{eq1e}
\end{equation}
to the poor agent at location $n$.

If the state of an agent located at position $m$ is denoted by
$|m>$, the state of the agent at position $n$ as $<n|$, and the
rule of the interactions as $\Re$, then the amount of resources
transferred could be denoted as $<n| \Re |m>$. This quantity
equals (\ref{eq1e}), i.e.,
\begin{equation}
< n|\Re |m >  = (u_m  - u_0 ) \cdot \frac{{(u_0  - u_n ) \cdot
\theta (u_0  - u_n )}}{{\sum\limits_{i(m)} {(u_0  - u_i ) \cdot
\theta (u_0  - u_i )} }} \cdot \frac{{(u_m  - u_0 ) \cdot \theta
(u_m  - u_0 )}}{{\sum\limits_{j(n)} {(u_j  - u_0 ) \cdot \theta
(u_j  - u_0 )} }}, \label{eq1}
\end{equation}
which complies with the rule stated in the second section - agents
give part of their surplus, the transfer is local, the transfer
does not deteriorate the status of rich agents locally in time,
while it changes the status of agents lacking resources. The step
function $\theta(\alpha)$ equals $1 \;\; (0)$ for $\alpha>0 \;\;
(\alpha<0)$. The rule of interaction $\Re$  is a particular
realization of one value set. Among all value sets, a few of them
are proper for a certain social system. The construction $<n| \Re
|m>$ measures the strength of interaction conducted in accordance
with the set $\Re$. Expression (\ref{eq1}) is a formal counterpart
of the analysis of social constructions, like norms and rules of
which (\ref{eq1}) is an example, as an insurance against time and
local fluctuation of production \cite{M16}.

The sum of resources of every interacting pair of agents is
conserved in the interaction, in contrast to the agent-environment
interaction and the agent's internal dynamics.

The model is time-discrete. In this sense, the quantities $c$,
$a$, and the change in resources  $\Delta u$ in one time unit are
the rate of resource consumption, rate of average resources input
and rate of resources change, respectively. Therefore, a poor
agent has amount of resources $u$ smaller than the corresponding
consumption level $c$. Generally, time scales for agent-agent
interactions and agent-environment interactions are different.
Hence, making them equal represents a restriction.

Based on the above considerations about the resource transfer of
an agent at position $m$ between two subsequent moments $k$ and $k
+ 1$, the following relation holds:
\begin{equation}
u_m (k + 1) - u_m (k) =  - c + w(m,{}^{}k) + \sum\limits_{n(m)} {(
< m|\Re |n >  -  < n|\Re |m > )}. \label{eq2}
\end{equation}
In (\ref{eq2}), $w(m, k)$ is the value of the Gaussian random
variable in the $k$-th time unit evaluated at position $m$. During
simulations, resources are first reduced for $c$, then changed
because of $w$, and, finally, intra-agent contributions are
evaluated.

The model is described as belonging to a class of models with
manifestly local interaction. The interaction described is a
particular example of screened, short-range interaction. The
screening is realized through taking into considerations the
nearest neighbor agents. The range of interaction is related to
giving of resources only between the nearest neighboring rich and
poor pairs of agents. The presence of the widely accepted set of
rules means that there exist the global characteristics of a
system. Its universal acceptance among agents is a particular type
of interaction. As we do not explicitly consider the genesis of
the set of rules for agent dynamics, it is appropriate not to
treat it on an equal footing as microscopic dynamics. In other
words, the time scale on which the changes of $\Re$  develop and
evolve is considerably larger than the time interval in which the
system dynamics is determined.

The initial state of the system is that in which resources of all
$N = N_0^2$ agents equal $u_0$. The boundary conditions are
periodic, i.e. the agents at locations ($N_0, j$) and ($1, j$) are
first neighbors. This formal simplification is not substantial,
because the relative augmentation of the resulting number of
nearest neighbors is of the order of $1/N_0$.

Finally, let us briefly discuss the properties of our model in
relation to the rapidly developing field of complex networks. The
research of complex networks is focused on the networks of very
complicated structure and random character, see reviews
[\onlinecite{M17} - \onlinecite{M19}] and references therein. The
investigation of various topological characteristics of complex
networks is of significant importance for the understanding of
numerous real and vital networks, such as the Internet, WWW and
many others. The structure of the network describing intra-agent
interactions in our model is fairly simple and regular. However,
there is no conceptual obstacle for the implementation of our
model's dynamics on the system of agents situated at the nodes of
some more complex network, e.g. scale-free network. Such, more
profiled modeling would deepen the insight into both the social
dynamics and the structure and dynamics of the complex networks.
The latter is realized in at least two modes. First, our model
introduces the thermodynamic description of the network underlying
social dynamics, thus its extensions contribute to the development
of the thermodynamically inspired description of complex networks.
Secondly, the interaction rules are the formalization of
collective attempt at preserving the integrity of the network
underlying social system in a stochastic environment. One could
argue that by developing the last point one gets the operationally
valuable collective mechanisms for the maintaining of the network
functionality in the uncertain (e.g., stochastic) environment.

\section{Indicators}

\subsection{Indicator set}

States of the model are generally, physically non-stationary
states. However, in a special case of $a = c$, the resources
average net transfer is zero, hence an almost stationary resource
flow of intensity $a$. Non-stationarity is then a consequence of a
variable number of agents. When, furthermore, such a change is
relatively small, a virtually stationary situation occurs.

Indicators attributed to a system state differ in origin. One set
of them originates in physics and includes, e.g., physical free
energy $F$, which is considered here in detail, entropy $S$,
temperature denoted here as $T$. Other indicators are more similar
to social indicators: number of agents $N$, and surplus of
resources. The formulas for indicator determination are written
having in mind restrictions of their validity induced by
non-stationarity.

Entropy is calculated using maximum entropy principle, thus
\begin{equation}
S =  - N\int\limits_0^\infty  {p(u)\ln p(u)du}, \label{eq3}
\end{equation}
where $p(u)$ is numerically determined distribution of agent
resources. It is taken that (\ref{eq3}) gives the values of both
physical and social entropy. That is not always valid  \cite{M20}.
Here it is a consequence of only one type of resources and the
measure associated with it. In more complex models, several types
of resources are explicitly treated, hence the need to
differentiate e.g., material and information flows  \cite{M20}.
Furthermore, expression (\ref{eq3}) is developed within the
equilibrium statistical physics. A seemingly more proper way to
calculate entropy would be to use the principles appropriate for
stationary states, like minimal entropy production or maximum
power production. However, in general, more realistic adoption of
these principles is to attribute different value sets to different
classes of agents, thus describing a part of agents using minimal
entropy production principle, another part of agents using the
maximum power production, and the rest of the agents using some
other principle(s). Because of that, the use of a single
extremization principle is by no means more correct than the use
of (\ref{eq3}) to calculate $S$. Therefore, the determination of
entropy needs to be prescribed in the least presumptuous, yet
objective way \cite{M21}. These conditions are fulfilled with
(\ref{eq3}).

The temperature is generally defined as
\begin{equation}
T = \left( {\frac{{\partial U}}{{\partial S}}}
\right)_{V,{}^{}N,{}^{}q},\label{eq4}
\end{equation}
in which $V$, $N$, $q$ are constant space, number of agents and
the flow from the environment to a system. Here, the temperature
is calculated during the system evolution as
\begin{equation}
T = \left( {\frac{{\partial U}}{{\partial S}}} \right)_V .
\label{eq4a}
\end{equation}

The internal energy is the sum of individual agent resources
\begin{equation}
U = \sum\limits_i {u_i }, \label{eq5}
\end{equation}
and for this model it is the Lyapunov function, as its time
derivative satisfies
\begin{equation}
\dot{U} = N(a-c), \label{eq6}
\end{equation}
from which the asymptotical character of the system state is
deduced.

The indicators introduced up to this point are auxiliary, in the
sense that they enable the reader to understand the model dynamics
in more detail. The indicators relevant for the objective of the
paper are the following: the physical free energy of a system,
which is given by
\begin{equation}
F = U - TS, \label{eq7}
\end{equation}
and determined by using (\ref{eq3}), (\ref{eq4a}) and (\ref{eq5});
the surplus
\begin{equation}
F_s  = \sum\limits_m {\theta (u_m  - u_0 )(u_m  - u_0 )},
\label{eq8}
\end{equation}
which we call social free energy. The social free energy
(\ref{eq8}) is the amount of resources that the agents could
disseminate in accordance with (\ref{eq1}).

\subsection{Dynamics of auxiliary indicators}

The combination $a/c$ of the parameters of the model represents
the main part of the model dynamics. In cases of $a$ differing
significantly from $c$ the dynamics gets simplified into either a
rapid flourishing or a rapid collapse of a system. Then the very
existence of a system becomes questionable. The latency of the
model is not clearly represented, and it is more proper to
interpret the model as a representation of a transient structure.
Therefore, further in the text we concentrate on the case $a
\approx c$. The corresponding model states resemble stationary
states and the equations (\ref{eq3}, \ref{eq4a}, \ref{eq5},
\ref{eq7}) are appropriate. Furthermore, the system adaptation is
maximal, because there are no unused environment resources which
exist for $a < c$, while the efficiency of use of obtained
resources is not maximal in the case $a > c$. Aditionally, the
level of consumption $c$ is considered equal to the reference
level $u_0$. The model dynamics is simulated during $100$ time
units from the initial moment.

\begin{figure}
\includegraphics{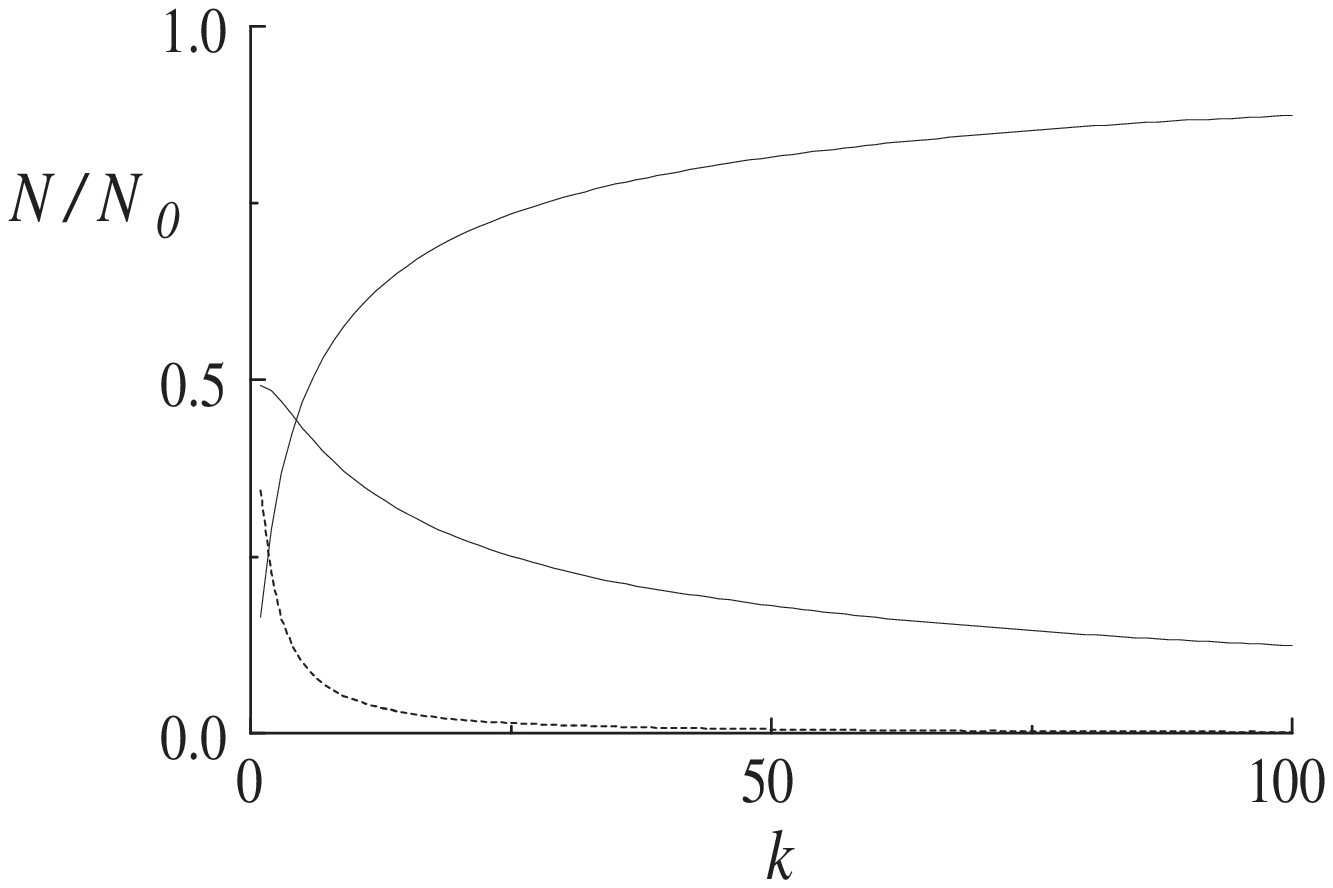}
\caption{\label{fig:f2} }
\end{figure}

In Fig.~\ref{fig:f2} the time dependence of the number of rich,
poor, and dead agents is given for $a/c = 0.9$. It is clear that
the changes in the number of live agents become negligible after
several time units. Then the system is balanced in the sense that
the influence of the initial state ceased, and the gradual
collapse of the system is not clearly seen.

\begin{figure}
\centering
\includegraphics[width=5.4in, height=4.5in]{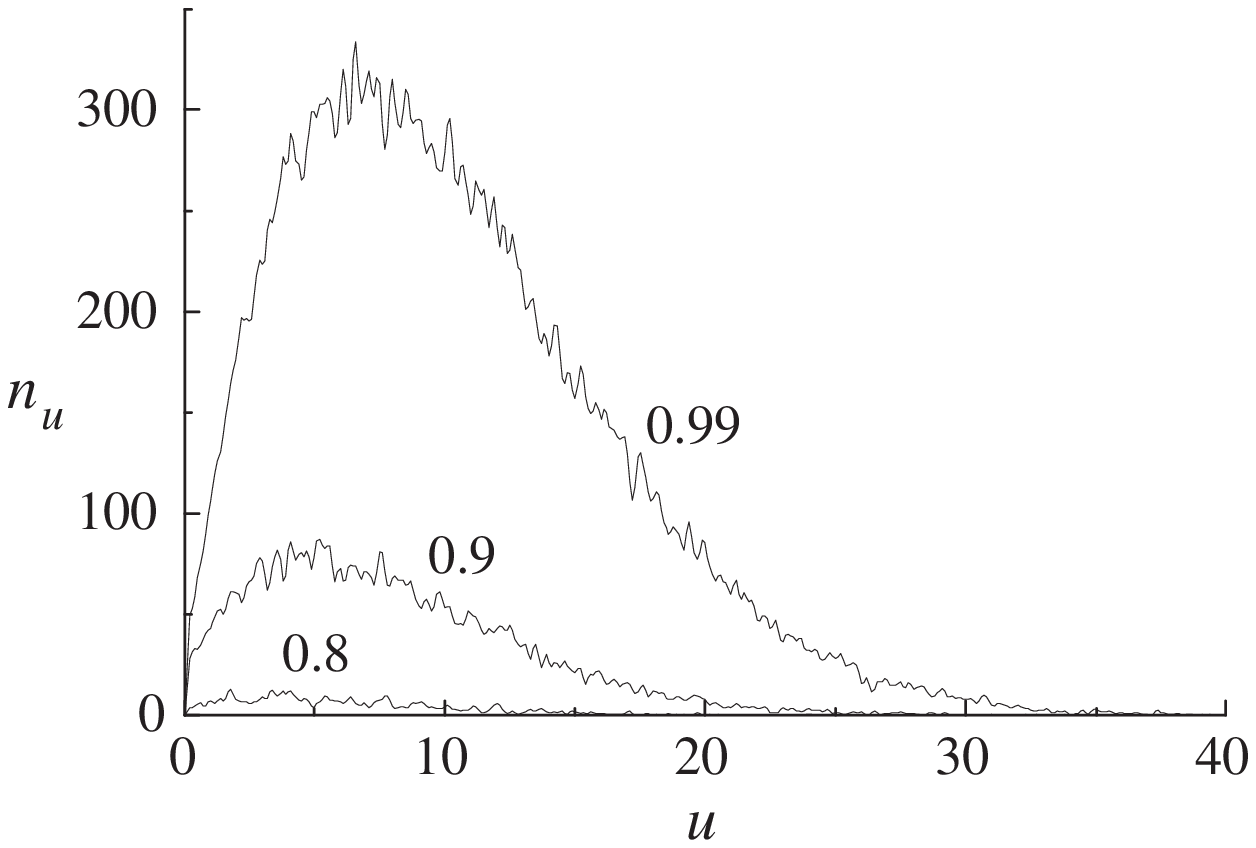}
\caption{\label{fig:f3} }
\end{figure}

The distribution of resources among agents is shown in
Fig.~\ref{fig:f3}. All graphs shown contain one maximum and a
localized tail on the side of high resources.

For large enough $a/c$, the temperature formally attains a
negative value at the beginning. However, that cannot be readily
interpreted as negative thermodynamic temperature as the system is
then in an intensively non-equilibrium state and the very
applicability of (\ref{eq4a}) is questionable, similarly to the
questionable applicability of other physical formulas.

\section{Determination of free energies}

\begin{figure}
\includegraphics{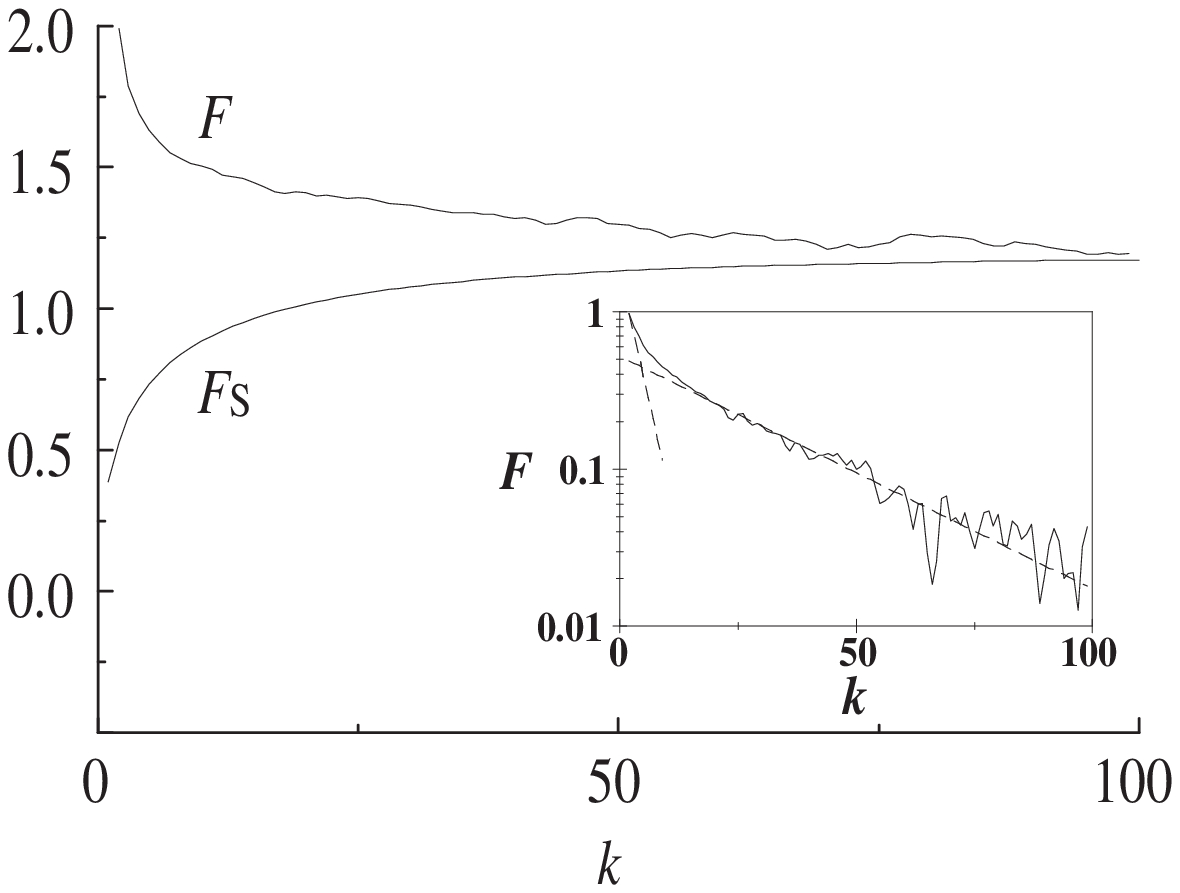}
\caption{\label{fig:f4} }
\end{figure}

The time dependence of physical and social free energies is shown
in Fig.~\ref{fig:f4}. Physical free energy is fitted to a double
exponential decay
\begin{equation}
F = C_1 \exp ( - t/t_1 ) + C_2 \exp ( - t/t_2 ), \label{eq9}
\end{equation}
where the dependence of the parameters $C_{1,2}$ on $a/c$ is
suppressed. In the inlet of Fig.~\ref{fig:f4}, lines representing
two decaying contributions to physical free energy are explicitly
shown for $a/c = 0.9$. The ensemble averaging does not change
significantly the results, which are presented non-averaged. Time
$t_2$ in (\ref{eq9}) diverges for $a \rightarrow c$ as described
with the following form
\begin{equation}
t_2 = \frac{\tau}{(1-a/c)^\kappa}
\end{equation}
in which  $\kappa = 1.29 \pm 0.01$  and  $\tau = 1.8 \pm 0.1$.

\begin{figure}
\includegraphics{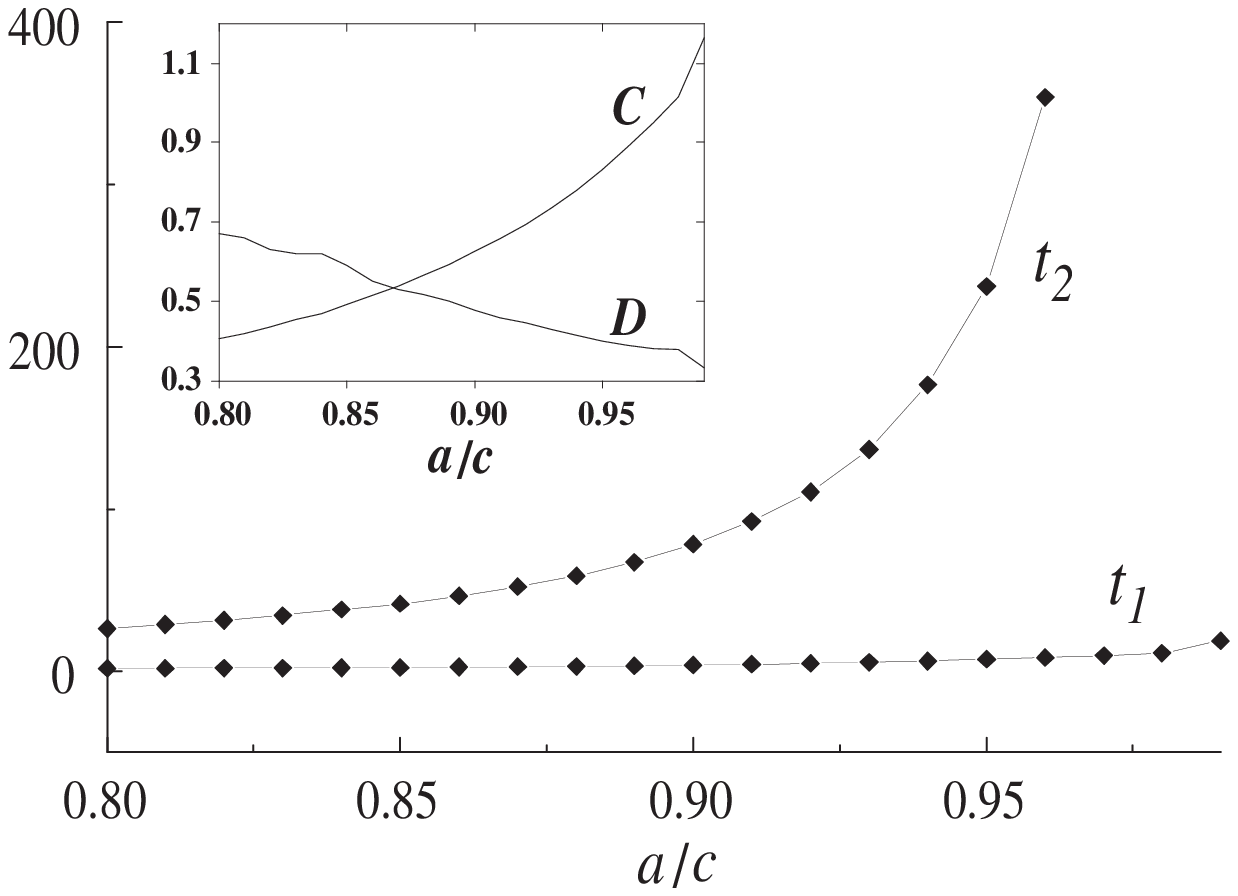}
\caption{\label{fig:f5} }
\end{figure}

\begin{figure}
\includegraphics{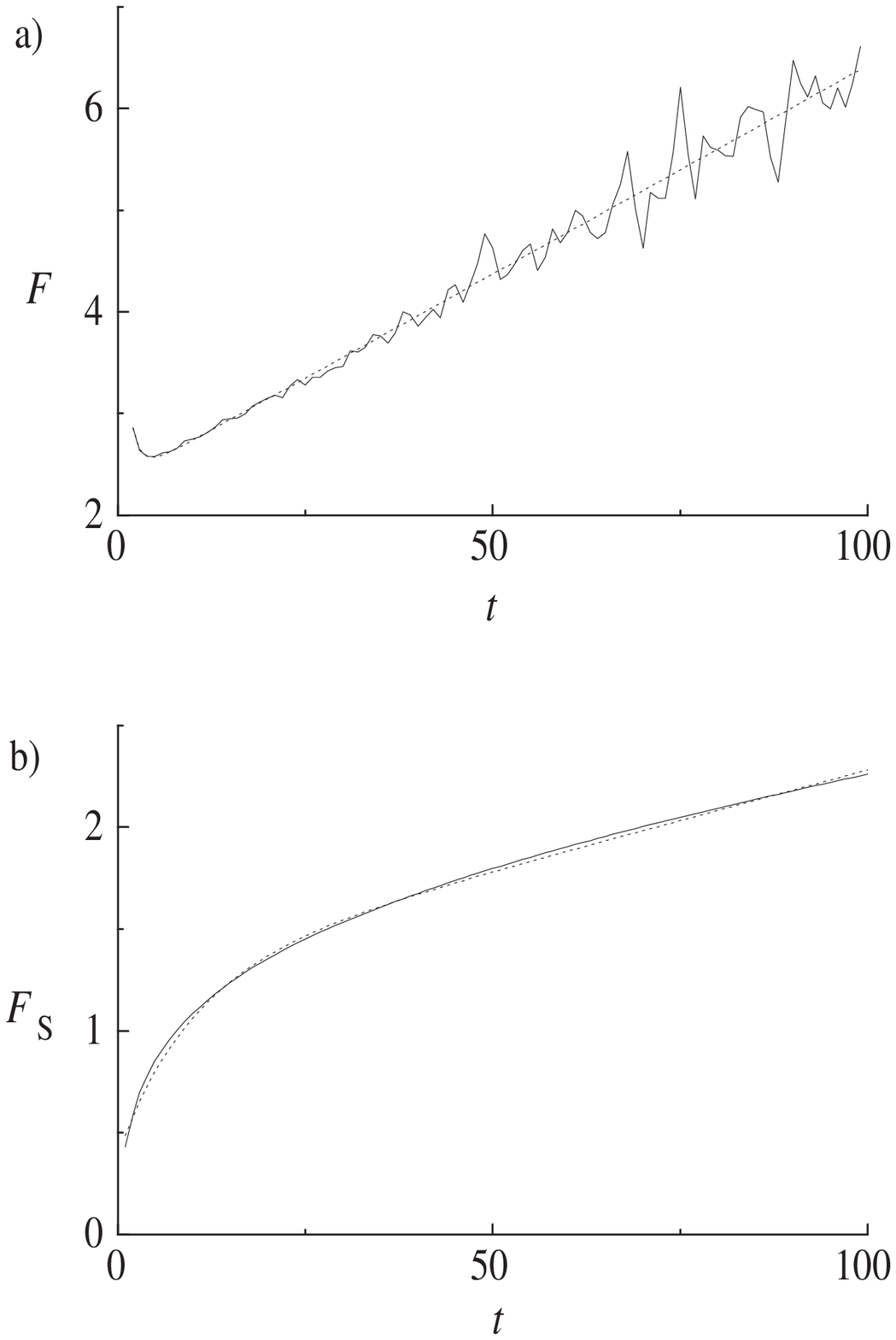}
\caption{\label{fig:f6} }
\end{figure}

The social free energy is fitted to the impulse function
\begin{equation}
F_s = C (1 - \exp ( - t/t_1 ) )^D  \exp ( - t/t_2 ), \label{eq10}
\end{equation}
with $C$, $D$ and $t_{1,2}$ depending on $a/c$. Fig.~\ref{fig:f5}
shows the dependence of parameters in (\ref{eq10}) on $a/c$. Time
parameter $t_2$ in (\ref{eq10}) diverges for $a/c \rightarrow 1$,
similarly to $t_2$ in (\ref{eq9}). The typical form of free
energies for $a > c$ is given in Fig.~\ref{fig:f6}, with the
fitting function form
\begin{equation}
F = C_1 \exp ( - t/t_1 ) + C_0 + M \cdot t , \label{eq11}
\end{equation}
valid for both physical and social free energy. The factor $C_1$
for the physical free energy fit has the same meaning as in
(\ref{eq9}). The numerical estimates for the coefficients in
(\ref{eq11}) relevant to short-time behavior, i.e., $C_1$ and
$t_1$, for the physical free energy have relatively large
deviations because they are influenced by large-time fluctuations.

\begin{figure}
\includegraphics{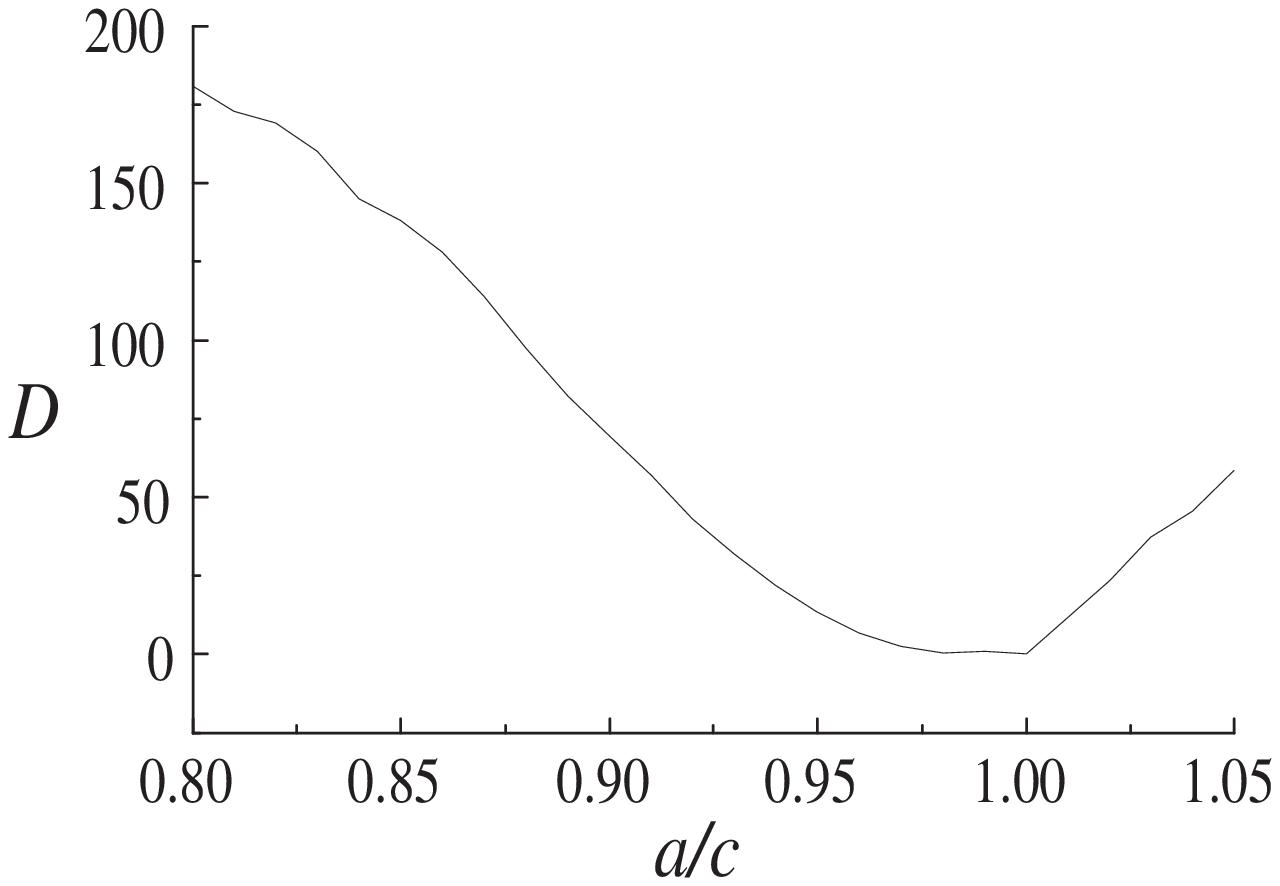}
\caption{\label{fig:f7} }
\end{figure}

One can express the difference between the fitting functions for
physical and social free energy by integrating the squared
relative difference of these two functions in the time interval in
which the form (\ref{eq4a}) is applicable. Since there is no
preferred function between them, their difference is compared with
their arithmetic mean in obtaining the relative value. The
difference function is taken as
\begin{equation}
D(a/c) \equiv \int\limits_{30}^{80} {\left| {\frac{{F_s (t) -
F(t)}}{{[F_s (t) + F(t)]/2}}} \right|^2 dt}. \label{eq12}
\end{equation}
Its dependence on $a/c$ is shown in Fig.~\ref{fig:f7}. The
conditions in (\ref{eq12}) are that relaxation of initial state
and long-time dynamics are excluded from the integration range,
which is why it is restricted from $t = 30$ to $t = 80$.
Relatively small changes of $D$, caused by small changes of
integration limits, are therefore admissible.

The two different decay times in expression for physical free
energy (\ref{eq9}) are connected with two different processes.
Faster decay is connected with rapid dying of agents whose initial
exchange of resources is negative and absolutely larger than $c$.
The time constant $t_1$ represents the memory duration of the
model in the sense that the influence of the initial state becomes
negligible. These results point to the fact that the dynamics of
the initially micro-canonical distribution coupled to the
stochastic environment is considered. Asymptotically and for $a <
c$, in time of the order of $t_2$ the system gradually collapses,
its number of agents and free energies tend to zero. In times $t$
larger than several $t_1$ and smaller than several $t_2$ the
system for $a \approx c$ is approximately a closed system. It is
this time interval for which the equilibrium form of physical free
energy (\ref{eq6}) can be reliably used, because then the
non-stationarity of the resource flow is relatively small and the
number of live agents is relatively constant, Fig.~\ref{fig:f2}.
In these cases, there is significant similarity in values and
character of physical and social free energies, Fig.~\ref{fig:f4},
despite the fact that their functional forms are different, as
seen from (\ref{eq9}) and (\ref{eq10}). It should be pointed out
that these functional forms are the consequences of rather
different starting points: physical free energy is introduced
using standard physical formalism, which is independent of a
model, thus generally valid as far as its states are
quasi-equilibrium states. On the contrary, social free energy is
introduced as a socially rather intuitive quantity - a surplus of
resources. It is the quantity defined for this particular model.
Yet, these two quantities are functionally and quantitatively
similar in a class of quasi-stationary states of a model.

The minimum of the relative difference between the free energies,
$D$, attained for $a = c$ contributes to the statement that $F$
and $F_S$ are equivalent. In case $a = c$ the system behavior is
expected to be the closest to the equilibrium one.
Fig.~\ref{fig:f7} shows in a more precise form that the alignment
between the $F$ and $F_S$ is the largest in the case in which the
equilibrium physics approach has the largest applicability. The
same functional form for both free energies in case of $a > c$ is
a consequence of the gained stationarity of states in the sense
that the number of agents for $a/c > 1$ virtually does not change
after several $t_1$ passed. The contribution $M t$ in that case,
as figuring in (\ref{eq11}) is a consequence of the net input of
resources from the environment.

In a more developed model, in which there are explicit mechanisms
for changes of the values of the defined parameters, the
purposefulness of a system development could be introduced. Then
the transfer of additional resources related to other purposes
could be defined. Such transfers could contribute to internal
system development, relatively independently of the environment.

\section{Summary and conclusions}

In this paper, the emphasis is put on the relation between the
social and physical free energies. Their equivalence for a class
of quasi-stationary model states is shown. The free energy in this
model has a clear meaning of surplus of resources. Despite the
relatively restricted class of states for which the equivalence of
the two free energies is shown, because of the different time
dependence of their fitting functions, it is conjectured that
physical and social free energy are different representations of
the same function. This is to be emphasized as physical free
energy is defined within the model-free formalism, while the
social free energy is an intuitive measure of surplus of
resources. Overall, the results obtained give preliminary insight
into the meaning of social free energy, and the class of system
states for which the social free energy is equivalent, or at least
similar to the physical free energy. On the one hand, further
analyses of more realistic models are needed in order to make that
relation clearer. On the other hand, introduction of free energy
followed by its interpretation within the social context raises a
number of further questions regarding the social interpretation of
different concepts of equilibrium and non-equilibrium physics.

Furthermore, in follow-up work on this model more profiled forms
of thermodynamic functions, e.g., Gibbs energy, are to be used in
order to incorporate a variable number of agents. In addition, the
intra-system generation of new agents is to be included. In this
case, the truly stationary states are possible, bringing about the
possibility of testing the equivalence of free energies in a
broader class of states. The structure of the net is rather
simplified, hence the inclusion of realistic, more complex
structure of nets is needed.

\begin{acknowledgments}
The authors acknowledge the fruitful discussions with Z. Grgic.
\end{acknowledgments}

\newpage

\section*{List of figures:}

\begin{itemize}
\item Figure 1. Two-dimensional net with agents. Two of the
agents, $A_{ij}$ and $A_{pq}$, are emphasized in order to explain
the principle of the agent-agent interaction. To determine the
total amount of resources that the rich agent will give to the
poor one, the total resources of their nearest neighborhoods are
considered. Circles denote the nearest neighbors of agent
$A_{ij}$.

\item Figure 2. Time dependence of a number of agents in the
system, for $a/c = 0.9$. Dashed line - number of poor agents. Full
lines denote the number of dead (rise in time) and live (fall in
time) agents. The initial number of agents is $N_0 = 40000$.

\item Figure 3. Distribution of resources among agents in time
unit $k = 100$. Numbers in the graph are values of $a/c$.

\item Figure 4. Time dependence of thermodynamic free energy $F$
and social free energy $F_s$ for $a/c = 0.99$. Inlet: separate
contributions to the double exponential fit of $F(k)$ for $a/c =
0.9$ shown in the log-linear plot. Full curve is thermodynamic
free energy. Dashed lines are $\log F = \log (0.8642)-k/3.268$,
and $\log F = \log(0.523)-k/29.32$ as fast and slow decaying
component, respectively.

\item Figure 5. Dependence of the characteristic times in fit
(\ref{eq10}) of social free energy on $a/c$. Inlet: dependence of
$C$, and $D$ on $a/c$.

\item Figure 6. Results (full lines) and fitting functions (dashed
lines) (\ref{eq11}) for $a/c > 1$. a) Thermodynamic free energy
with $C_0 = 2.30 \pm  0.08$; $C_1 = 2 \pm  2$; $t_1 = 1.5  \pm
1.3$ and $m = 0.0423 \pm 0.0035$, b) social free energy with $C_0
= 1.29  \pm 0.01$; $C_1 = - 0.90 \pm 0.02$; $t_1 = 9.9  \pm 0,5$
and $m = 0.0099 \pm 0.0002$.

\item Figure 7. Dependence of the measure (\ref{eq12}) of
difference between thermodynamic and social free energy on $a/c$.

\end{itemize}

\newpage

\newpage

\newpage

\newpage

\newpage

\newpage

\newpage

\end{document}